\documentclass[a4paper,11pt]{article}
\pdfoutput=1 

\usepackage{jinstpub} 

\usepackage{graphicx}
\usepackage{bm}
\usepackage{amsmath}
\usepackage{multirow}
\usepackage{stackrel}
\usepackage{float}
\usepackage{subcaption}

\title{Sparse Deconvolution Methods for Online Energy Estimation in Calorimeters Operating in High Luminosity Conditions}

\author[a]{Tiago Teixeira}
\author[a]{Luciano Andrade}
\author[b]{and Jos{\'e} Manoel de Seixas}
\affiliation[a]{Electrical Engineering, Federal University of Juiz de Fora,\\Minas Gerais, Brazil}
\affiliation[b]{Signal Processing Lab, COPPE/POLI - Federal University of Rio de Janeiro,\\Rio de Janeiro, Brazil}

\emailAdd{tiago.teixeira@engenharia.ufjf.br}

\abstract{Energy reconstruction in calorimeters operating in high luminosity particle colliders has become a remarkable challenge. In this scenario, pulses from a calorimeter front-end output overlap each other (pile-up effect), compromising the energy estimation procedure when no preprocessing for signal disentanglement is accomplished. Recently, methods based on signal deconvolution have been proposed for both online and offline reconstructions. For online processing, constraints concerning fast processing, memory requirements, and cost implementation limit the overall performance. Offline reconstruction allows the use of Sparse Representation theory to implement sophisticated Iterative Deconvolution methods. This paper presents Iterative Deconvolution methods based on Sparse Representation algorithms whose computational cost is effective for online implementation. Using simulated data, current techniques were compared to the proposed Sparse Representation ones for performance validation in the online environments. Analysis has shown that, despite the higher computational cost, when compared to standard methods, the performance improvement may justify the use of the proposed techniques, in particular for the Separable Surrogate Functional, which is shown to be feasible for implementation in modern FPGAs.}


\begin{document}
\maketitle
\flushbottom

\section{Introduction}
High-Energy Physics (HEP) experiments usually require sophisticated instrumentation systems, especially when particle colliders are involved~\cite{Barletta2014}. In order to analyze increasingly rare physical phenomena, there is a tendency of increasing their collision rates and the number of particles per collision, thus increasing the luminosity achieved by the experiment~\cite{Bruning2015}. This is the case of the most potent collider ever built, the LHC (Large Hadron Collider)~\cite{Evans2008} at CERN (\textit{Organisation Européenne pour la Recherche Nucléaire})~\cite{Anthony2014}. The LHC collides bunches of protons every 25 nanoseconds and has been establishing records in terms of both collider energy and luminosity~\cite{Pralavorio2017}. Moreover, upgrades that are on the way for the next LHC operation periods (Run 3 and beyond) will increase even more the luminosity~\cite{Cepeda2018}, towards a projected peak around 5 to 7.5 $\times 10^{34}cm^{-2}s^{-1}$~\cite{Walkowiak2018}.

For new physics identification at the LHC, calorimeters~\cite{Wigmans2018} play an important role, as they measure the energy of the incoming particles and are essential for triggering purposes~\cite{Jeitler2017}.  The two general-purpose LHC experiments (ATLAS~\cite{Aad2008} and CMS~\cite{Chatrchyan2008}) comprise  hundred of thousands of calorimeter readout channels each and employ state-of-art detector technology. However, the LHC's enormous luminosity imposes severe restrictions on calorimetry performance as, despite being fast, the signal generated in the detector's front-end electronics requires a certain number of bunches crossings (BCs) to be fully developed. Therefore, signal pile-up effects may arise when the luminosity is increased.
 

The trigger system is often implemented through an online cascade signal processing chain consisting of two or more levels, where the rate reduction is proportional to the level of the algorithm complexity for final event selection. Due to the high event rate, the lower filtering levels are usually implemented in hardware and rely very much on online energy estimation, which is addressed in this paper for applications where signal pile-up becomes an issue.


\subsection{The Pile-up Effect}

\begin{figure}[htb]
\centerline{\includegraphics[width=.6\textwidth]{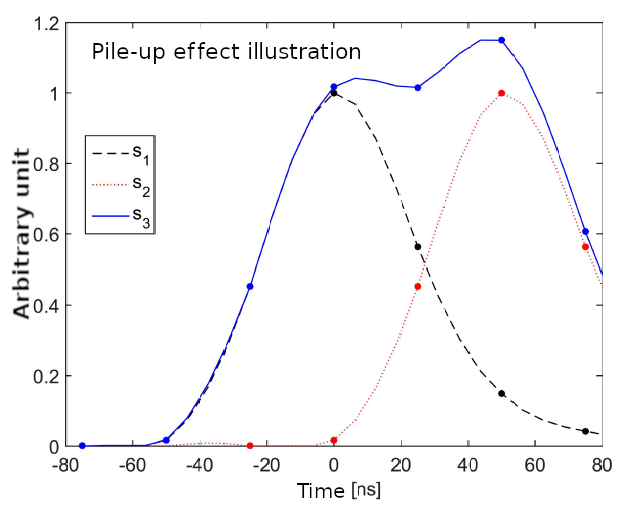}} 
\caption{Example of the pile-up effect. See text.}
\label{fig:pile-up}
\end{figure} 

An example of out-of-time signal pile-up in collider experiments can be seen in Figure~\ref{fig:pile-up}, where signals were digitized by a 40 MHz clock ($25ns$ period, as in the LHC). The unipolar signal generated in a hypothetical calorimeter cell by a given collision is referred to as $s_1$, and $50ns$ later a new signal ($s_2$) from two bunch collisions after reaches that particular cell. Supposing a $150 ns$ readout acquisition length, the $s_3$ signal is formed, which is actually the stacking of both $s_1$ and $s_2$ signals. This pile-up effect leads to a sub-optimal operation of the classical energy estimation algorithms, which are based on Optimal Filtering (OF) theory~\cite{Fullana2006, Radeka1988, Bertuccio1992, Cleland1994}.

The calorimeter signal shape is a linear combination of a signal with neighboring pileup signals, where the signal amplitude is proportional to the target energy value~\cite{Wigmans2018}. Therefore, the sensor output can be modeled as the convolution between the target sequence (energy deposition per bunch crossing) and the calorimeter normalized (unitary amplitude) reference pulse shape. To recover the target sequence, a deconvolution algorithm may be applied~\cite{Haykin2013}.

Different deconvolution algorithms have been proposed for energy estimation in high-luminosity calorimeter operation. The work in~\cite{AndradeFilho2015} proposes an offline procedure in two steps, where a signal detection per bunch crossing is performed first, followed by a multi-amplitude estimation procedure. Still focusing on offline reconstruction, the work in~\cite{Barbosa2017} implements an approach based on Sparse Representation (SR)~\cite{Elad2010} of data, presenting remarkable results, over-performing standard deconvolution methods. Sparse Representation was originally developed and applied in signal compression and image denoising routines~\cite{Silva2018, Starck2010}. The work in~\cite{Barbosa2017} inaugurated the Sparse Representation energy reconstruction applications. However, its cost implementation is presently forbidden for online data-stream.



Concerning online processing, a deconvolution method based on Finite Impulse Response (FIR)~\cite{Mitra2010} filters was proposed in~\cite{Duarte2019} and it has shown to be very attractive due to its lower cost implementation in hardware. However, recently Sparse Representation theory has developed families of algorithms targeting cost implementation reduction~\cite{Lim2018, Zaki2018, Niu2019}. Although FIR filter structures are known to be the most economic approach in terms of hardware resources, the cost-effectiveness of such modern Sparse Representation techniques has paved the way for online energy reconstruction implementation. Those approaches have already been used in several online applications like machine learning~\cite{Ma2017} and big data~\cite{Zhao2020}.

Therefore, the main goal of this paper is to use modern Sparse Representation theory in a design that fulfills online requirements. The adaptation of some of those techniques is proposed envisaging energy estimation in calorimeter cells, opening a new research area in online energy reconstruction. Since Sparse Representation has shown to be the most accurate energy reconstruction procedure under pile-up, the methods proposed here can be used as an alternative for FIR filter implementation in applications where the hardware resource minimization is not the main constraint. A comparison among the aforementioned methods, in terms of cost implementation and energy reconstruction performance, is also provided, indicating the best situation for the usage of each approach.

In order to complete the discussion, a possible implementation for one of the proposed Sparse Representation methods (in this case, the Separable Surrogate Functional) is outlined at the end of the paper. The circuitry was developed in Hardware Description Language (HDL)~\cite{Botros2015} envisaging a straightforward implementation in Field-Programmable Gate Array (FPGA) technology~\cite{Meyer-Baese2007}.

\section{Techniques}

In this section, firstly the FIR approach for high-luminosity online energy estimation is briefly reviewed, as it is being used here for baseline comparison.  In the sequence,  the basics of modern sparse signal processing are presented, as a way to exploit the nonlinear statistics that comes with the out-of-time signal pileup.

\subsection{FIR Filter Based Energy Estimation}

\begin{figure}[htb]
\centering
\includegraphics[width=.95\textwidth]{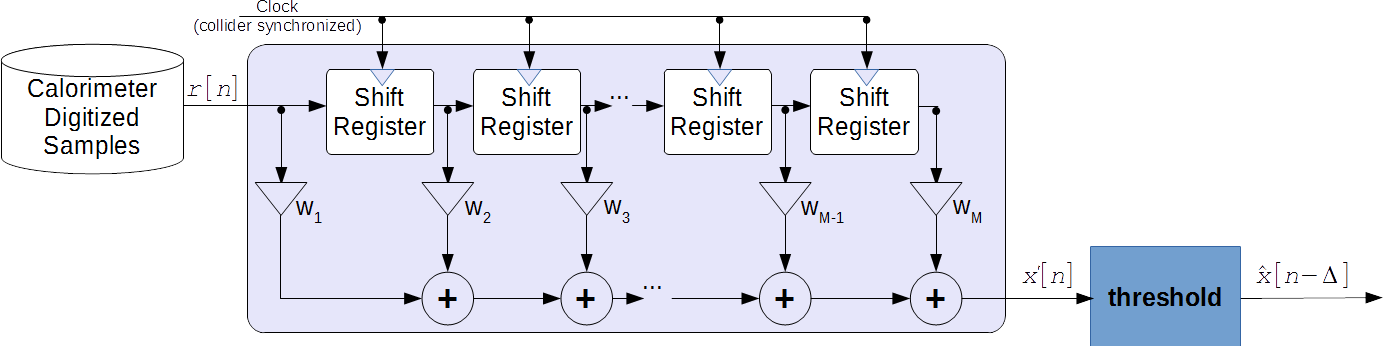}
\caption{The FIR filter structure implementing signal deconvolution in online energy estimation, based on~\protect\cite{Duarte2019}.}
\label{fig:fir}
\end{figure}

Figure~\ref{fig:fir} shows the basic structure for online energy estimation based on a FIR deconvolutional filter~\cite{Duarte2019}. The signal $r[n]$ represents the measured information from a single front-end calorimeter sensor, after applying the free-running (an uninterrupted) Analog-to-Digital Converter (ADC). A cascade of registers, synchronous with the ADC clock, implements a shift-register structure~\cite{Meyer-Baese2007}, which is responsible for storing the latest $M$ digitized samples. The filter order ($M-1$) depends on both the calorimeter reference pulse-shape width and the signal pile-up intensity. For deconvolution, the filter order is often higher then the reference pulse width (two times on average), since the filter needs tail information of neighbor superimposed signals to perform the desired source separation. The equation employed in this deconvolution procedure is: 

\begin{equation}
\label{eq:fir}
x'[n]=\stackrel[l=0]{M-1}{\sum}(w[n-l]r[n])
\end{equation}

The weighting vector $\mathbf{w}$ is obtained offline, based on deconvolution
principles. Details on the filter design can be found in the original paper~\cite{Duarte2019}. The samples are, then, linearly combined at the filter output, which delivers new information at each clock. A threshold is applied to the filter output to avoid small artifacts from the deconvolution process in adjacent bunch crossings, mainly small negative energy estimations.

The system output corresponds to the reconstructed target information sequence $\hat{x}[n]$, which is a straightforward estimation of the energy deposit per bunch crossing ($x[n]$). The system starts delivering the output sequence $\Delta$ clock cycles after the first sample goes through the processing chain. This delay is necessary because the system needs information from previous samples in order to perform the deconvolution procedure. Typical values of $\Delta$ are around half of the filter order.

Despite of its very suitable hardware implementation, the purely linear operation of FIR filters does not allow access to the Higher Order Statistics (HOS)~\cite{Arce2005} present in the embedded pile-up noise, which may be exploited for further performance improvement. The proposed Sparse Representation based deconvolution methods access the Higher Order Statistics information indirectly through the usage of non-linear functions in the deconvolution process.

\subsection{The Sparse Representation}

The Sparse Representation theory has been applied for energy reconstruction with great success for offline environments in~\cite{Barbosa2017}. On that paper, the implementation was based on Linear Programming (LP)~\cite{Luenberger2008}, which is a standard procedure with high computational cost. Recently, modern Sparse Theory has raised with more cost-effective algorithms~\cite{Elad2010}. Modifications on standard Matching Pursuit (MP) methods~\cite{Mallat1993} have presented an intrinsic relationship with energy estimation tasks. In those approaches, a signal detection iterative procedure is executed first, using a bank of linear Matching-Filter (MF)~\cite{Hadley2010} based filters. After some iterations, all the superimposed signals within an acquisition window are detected. Their amplitudes are, then, estimated simultaneously using the Least Square (LS) algorithm~\cite{Kay1993}.

Another Sparse Representation family of methods that is being employed due to its excellent cost-benefit properties is the Iterative-Shrinkage (IS) algorithms~\cite{Chen2017}. These methods combine standard linear Coordinate-Descent (CD) algorithms~\cite{Takac2014} with a linear-by-part Shrinkage function, in order to promote sparsity, and becoming one of the most cost-effective algorithms for Sparse Representation.





\section{Proposed Sparse Representation Based on Deconvolution Methods}

The Sparse Representation based deconvolution methods are designed by using a windowed acquisition point of view. The method is guided by the matrix formulation from the convolution operation of finite size sequences, as proposed in~\cite{Barbosa2017}. Given a normalized calorimeter reference pulse-shape, represented by the vector~$\mathbf{h}$ of size~$V$ and a target sequence, represented by the vector~$\mathbf{x}$ (energy per bunch crossing) of size~$G$, the convolution between these two signals produces the measured signal $\mathbf{r}$; that is, the calorimeter readout channel response. Because there are $G$ bunch crossings, and the fact that the calorimeter responses extends $V$ samples beyond the last bunch crossing, the total samples considered for the vector $\mathbf{r}$ are $G+V-1$~\cite{Mitra2010}.  This convolution process can be written as:

\begin{equation}
\label{eq:matrix}
\mathbf{r}=\mathbf{H}\mathbf{x}
\end{equation}
where $\mathbf{H}$ is the convolution matrix of size $U \times G$ (columns of $\mathbf{H}$ contain shifted versions of the vector~$\mathbf{h}$, with zero padding). The windowed deconvolution procedure consists of recovering the sequence $\mathbf{x}$ when $\mathbf{r}$ is measured and $\mathbf{H}$ is known. When the size of the vector~$\mathbf{r}$ is greater than the one of~$\mathbf{x}$, there are infinite solutions for this system of equations. In~\cite{Barbosa2017}, it was shown that, for deconvolution purposes, the sparsest representation of $\mathbf{x}$ is the correct solution when the target sequence is characterized by impulsive signals.

According to the Sparse Representation theory~\cite{Elad2010}, the sparsest solution for Equation~\ref{eq:matrix} can be found solving the optimization Problem~$P_{\ell}$ for $0 \leq \ell \leq 1$

\begin{equation}
\label{eq:pl}
(P_{\ell}):\underset{\mathbf{x}}{min}\left\Vert \mathbf{x}\right\Vert _{\ell}^{\ell}\:subjected\,to\:\mathbf{r}=\mathbf{Hx}
\end{equation}
where
\begin{equation}
\label{eq:norm}
\left\Vert \mathbf{x}\right\Vert _{\ell}^{\ell}=\underset{b}{\sum}\left|x_{b}\right|^{\ell}
\end{equation}	
is the $\ell$-norm of the vector~$\mathbf{x}$.

For $\ell = 1$, this results in a typical Linear Programming problem~\cite{Candes2007}. The use of Linear Programming for Sparse Representation aiming at energy estimation has been outlined in~\cite{Barbosa2017}, where this method was shown to outperform other windowed deconvolution methods. However, the focus of that paper was on offline energy reconstruction, since Linear Programming presents a very high computational cost, becoming usually prohibitive for online execution. In the sequence, two families of Sparse Representation algorithms are presented in order to solve the $P_{\ell}$ problem with different perspectives, both focusing on computational cost-effectiveness procedures.

\subsection{Matching Pursuit}

For $ \ell = 0 $, Equation~\ref{eq:norm} results in the number of non-zero elements in $\mathbf{x}$ and the Sparse Representation solution given by $P_0$ becomes evident. However, $P_0$ is not a mathematically tractable problem~\cite{Elad2010}. This is a combinatorial problem infeasible to be computationally implemented in practice.

To avoid the test of all combinatorial possibilities, Matching Pursuit seeks for similarities between columns of the matrix~$\mathbf{H}$ and the vector~$\mathbf{r}$, selecting the smallest number of columns that, when linearly combined, produce the vector~$\mathbf{r}$ added to a small residue vector~$\mathbf{d}$.

Given the support matrix~$\mathbf{H}_s$, which comprises the selected columns of~$\mathbf{H}$, the residue vector~$\mathbf{d}$ is

\begin{equation}
\mathbf{d} = \mathbf{H}_s \mathbf{x}_s - \mathbf{r}
\label{eq:residue}
\end{equation}
where the vector~$\mathbf{x}_s$ has the amplitudes for each signal in~$\mathbf{H}_s$ (the columns of~$\mathbf{H}_s$). The components in~$\mathbf{x}_s$ are chosen in order to minimize the square module of the residue:

\begin{equation}
\epsilon^2 = \mathbf{d}^T\mathbf{d} = |\mathbf{H}_s \mathbf{x}_s - \mathbf{r}|^2
\label{eq:energy}
\end{equation}
where $T$ indicates a vector transposition.

The vector~$\mathbf{x}_s$ that minimizes the Equation~\ref{eq:energy} is obtained by the Least Square equation

\begin{equation}
\mathbf{x}_s = (\mathbf{H}_s^T \mathbf{H}_s)^{-1} \mathbf{H}_s^T \mathbf{r}
\label{eq:ls}
\end{equation}

The Matching Pursuit algorithms select columns of the matrix~$\mathbf{H}$ with the highest potential of minimizing $\epsilon^2$~\cite{Wu2012}. Depending on the way this task is performed, a specific variation of the basic Matching Pursuit algorithm is applied. In this work, the Orthogonal Matching Pursuit (OMP)~\cite{Cai2011} and the Least Square Orthogonal Matching Pursuit (LS-OMP)~\cite{Dumitrescu2013} will be outlined.

\subsubsection{Orthogonal Matching Pursuit}

This method selects the column $\mathbf{h}_j$ of the matrix~$\mathbf{H}$ that best matches the current residue $\mathbf{d}$, minimizing the cost function

\begin{equation}
Q(x_j) = |x_j\mathbf{h}_j - \mathbf{d}|^2
\end{equation}
The value $x_j$ that minimizes this function is the Matching Pursuit equation
\begin{equation}
x_j = \frac{\mathbf{h}_j^T\mathbf{d}}{\mathbf{h}_j^T\mathbf{h}_j}
\label{eq:inner}
\end{equation}
This value is computed for all~$\mathbf{h}_j$ columns not yet inserted in the support matrix~$\mathbf{H}_s$. The column with the smallest~$Q(x_j)$ value is selected, the new residue vector is computed and a new iteration is performed until~$\epsilon$ becomes lower than a given threshold~$\epsilon_0$.  It can be shown~\cite{Dumitrescu2013} that all the columns of~$\mathbf{H}_s$ are orthogonal to the remaining residue~$\mathbf{d}$.

\subsubsection{Least Squares Orthogonal Matching Pursuit}

The LS-OMP method updates temporarily~$\mathbf{H}_s$ with the current column~$\mathbf{h}_j$ in test and computes~$\epsilon$ directly (Equation~\ref{eq:energy}). The temporary matrix~$\mathbf{H}_s$ that gives the smallest~$\epsilon$ value is kept. The process stops when the current residue is lower than~$\epsilon_0$ as well. Different from OMP, which computes a simple inner product (Matching Pursuit equation) to each column not yet inserted in~$\mathbf{H}_s$, LS-OMP computes the Least Square (Equation~\ref{eq:ls}) instead. Therefore, the computational cost of LS-OMP is higher than OMP.

\subsection{Iterative-Shrinkage}

Iterative-Shrinkage algorithms are based on the Problem~$P_1$. However, a relaxation is employed in order to allow a small residual square error~$\epsilon_0$ on the constraint.

\begin{equation}
\label{eq:p1}
(P_{1,\epsilon_0}):\underset{\mathbf{x}}{min}\left\Vert \mathbf{x}\right\Vert _{1}^{1}\:subjected\,to\:|\mathbf{H}\mathbf{x} - \mathbf{r}|^2<\epsilon_0
\end{equation}

The problem $P_{1,\epsilon_0}$ is a typical Linear Programming problem, as directly performed in~\cite{Barbosa2017} for offline signal reconstruction. However, focusing on cost-effectiveness, firstly, this problem is transformed to an optimization routine without
constraint by the use of a Lagrange multiplier~$\lambda$~\cite{Elad2010} 
\begin{equation}
\label{eq:p1a}
(P_{1,\lambda}):\underset{\mathbf{x}}{min}\left\Vert \mathbf{x}\right\Vert _{1}^{1} + \lambda|\mathbf{H}\mathbf{x} - \mathbf{r}|^2
\end{equation}
where~$\epsilon_0$ is absorbed by~$\lambda$. The Iterative-Shrinkage algorithms propose solving Problem~$P_{1,\lambda}$ by means of iterative Coordinate-Descent methods: given an initial guess vector~$\mathbf{x}^0$, the optimal $\mathbf{x}_{opt}$ value can be recursively inferred in a small number of steps. However, the use of standard Coordinate-Descent methods, like Newton-Raphson~\cite{Kelley2003}, becomes prohibitive for $P_{1,\lambda}$ because of the discontinuity of the $\ell$-norm for $\ell = 1$~\cite{Elad2010}.

\begin{figure}[htb]
\centerline{\includegraphics[width=.65\textwidth]{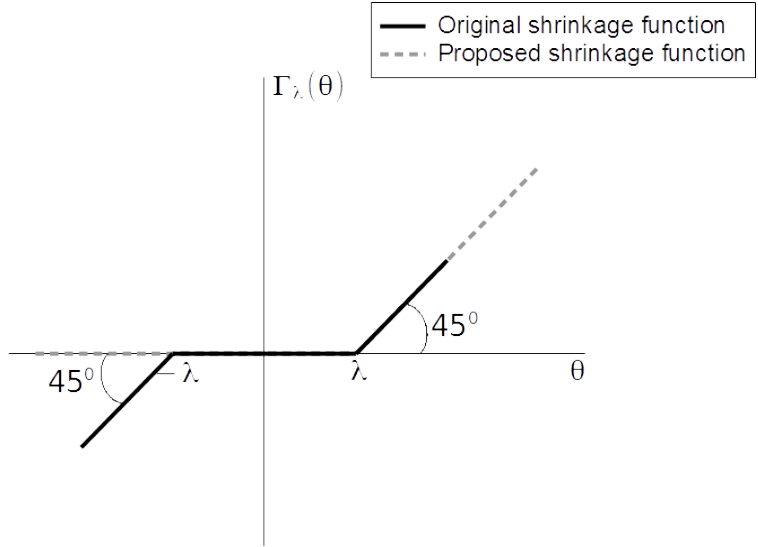}}
\caption{The original shrinkage function and the proposed modified version to promote positive energy estimation.}
\label{fig:shrink}
\end{figure}

In~\cite{Daubechies2004}, the insertion of terms in $P_{1,\lambda}$ is proposed, so that the minimum coordinate position~$\mathbf{x}_{opt}$ is not changed, but the multivariate problem is split into separated one-dimensional problems, which can be solved by parts. The resulting equation is named as Surrogate Function, and the respective Iterative-Shrinkage method is the Separable Surrogate Functional (SSF). After some algebraic manipulation, the iterative procedure can be compacted as follows

\begin{equation}
\mathbf{x}^{i+1} = \Gamma_{\lambda}\Bigg(\mathbf{x}^i + \mu[\mathbf{H}^T(\mathbf{r} - \mathbf{H}\mathbf{x}^i)]\Bigg)
\label{eq:shrink}
\end{equation}
where~$\mu$ is the step size in the direction of the minimum. The function~$\Gamma_{\lambda}(\theta)$ is the one-dimensional shrinkage function that should be applied to each component of the vector $\mathbf{x}$. In this paper, a modification on the shrinkage function is proposed in order to promote positive energy estimation, as shown in Figure~\ref{fig:shrink}. This approach increases the performance of the method for further iterations, since negative (and wrong) estimations are discarded along the process. With this modification, the resulting function can be simply implemented through a subtraction followed by a threshold~($\lambda$) operation. The argument in Equation~\ref{eq:shrink} is identified as a linear Gradient-Descent~(GD) iteration~\cite{Yun2010} and comprises only addition and multiplication operations. The parameters~$\mu$ and~$\lambda$ are both obtained from simulation.

\subsubsection{Separable Surrogate Functional Initialization}

One important prerequisite for online implementation of the Separable Surrogate Functional algorithm is to avoid a higher number of iterations. This can be accomplished by choosing a proper initial vector~$\mathbf{x}^0$. In this work, a preprocessing routine is proposed, in order to initialize the solution close to the optimal one. The proposal is to implement the Least Square (Equation~\ref{eq:ls}) for a full-support matrix~$\mathbf{H}$:

\begin{equation}
\mathbf{x}^{0}=(\mathbf{H}^{T}\mathbf{H})^{-1}\mathbf{H}^{T}\mathbf{r} = \mathbf{H}^{+}\mathbf{r}
\label{eq:x0}
\end{equation}
where $\mathbf{H}^{+}=(\mathbf{H}^{T}\mathbf{H})^{-1}\mathbf{H}^{T}$ is the pseudo-inverse of matrix~$\mathbf{H}^T$~\cite{Roman2008}. This is a constant matrix, which may be obtained offline.

\section{Performance Results}\label{sec:results}

In order to validate the proposed algorithms, a toy Monte Carlo simulation was used for generating a set of consecutive samples, which represent the target energy values, per bunch crossing, for a single readout channel. Energy distributions and sensor’s front-end were simulated following the characteristics found in modern calorimeters~\cite{Hrivnacova2003, Chapman2011, Banerjee2012}. The pile-up effect was assessed through the calorimeter cell occupancy: null occupancy corresponds to the absence of hits in a given run and on the other extreme, 100\% of occupancy means that an amount of energy is deposited at every particle collision that happens in a given run.

\subsection{Physical Model}

Concerning hardware implementation for the target online operation, Sparse Representation and FIR filters comprise very different approaches. The latter produces its output signal in a sample-by-sample sequence, which is very suitable and straightforward for free-running environments. On the other hand, Sparse Representation algorithms operate simultaneously on all the samples previously buffered within a window. Actually, this non-causal characteristics is one of the reasons of the increasing in performance, when compared to the FIR filter structures. Therefore, operating a windowed energy reconstruction algorithm in a free-running environment is one of the challenges addressed by the present work, as the online implementation of windowed methods demands extra digital circuitry for proper data-flow. In this paper, the proposed Sparse Representation methods are first demonstrated. Then, a hardware solution for online implementation is outlined.

Actually, the adaptation of such a non-causal method to online operation is possible thanks to some characteristics of modern colliders, as is the case of the LHC, which operates in a train-of-bunches~\cite{Shiltsev2014, Herr2005}. Due to particle injection stability issues, the accelerator needs small periods of empty-bunches intercalated with periods of bunch-trains~\cite{Fox2018}. At Run 2, for instance, typical physics runs were operated in a configuration of 48 filled bunches followed by 7 (or more) empty bunches~\cite{Hostettler2018}. It turns out that this issue encourages the use of windowed estimation methods as the ones proposed here, since such an operational approach eliminates two important design difficulties:

\begin{enumerate}
	\item \textbf{Window size determination} - in adapting windowed methods to free-running environments, the window size is to be determined. In a train-of-bunches configuration, it is natural to choose the window size in order to fit an entire train within the processing window.
	\item \textbf{Border effect} - A problem found in the windowing signal processing is that signals on the window border are split into neighbor windows. As a consequence, the estimation of bunch crossings closer to the window edge tends to reduce the processing efficiency. However, due to the train-of-bunches configuration, providing that the calorimeter reference signal width is smaller than the empty-bunches period, the border effect does not occur.
\end{enumerate}

In this paper, bunch-trains of 48 bunches followed by 7 empty bunches were used. Thus, window lengths of 55 bunches (48 central plus the guard comprising three bunches on each side of the window and an extra bunch between windows) were employed to avoid the border effect.

Without loss in generality, the cell calorimeter shaped pulse was considered as unipolar~\cite{Peralva2013, Sugiyama2015, Keil2007}, Gaussian-like, with an acquisition length of $150~ns$. The collision rate was of $40~MHz$ (as in the LHC), resulting in a seven-sample pulse representation where the pulse peak corresponded to the central sample ($4^{th}$ sample). It is worth mentioning that the proposed method might be applied to any pulse shape, provided it represents the impulse response ($h[n]$) to be deconvolved.


For optimal parameter determinations and for performance evaluations, the RMS of the error between the target ($x[n]$) and the estimated ($\hat{x}[n]$) sequences was employed (Equation~\ref{eq:rms})

\begin{equation}
RMS~Error = \sqrt{ \frac{ \sum_{n=1}^{N} (\hat{x}[n] - x[n])^2}{N} }
\label{eq:rms}
\end{equation}
where $N$  was the total number of samples in the test set (10,010).

\subsection{Parameters Choice}

The simulation framework employed here was also used in recent related works~\cite{Barbosa2017, Duarte2019}, where more details can be found. The signal characteristics were:
\begin{itemize}
    \item The cell occupancy range: from 1\% to 90\%, in steps of approximately 10\%.
    \item Energy deposition: from an exponential distribution of 360~MeV.
    \item Phase shift: an uniformly random pulse-phase in the range $\pm1$~ns is employed to each generated signal.
    \item Electronic noise: zero-mean white Gaussian noise with $\sigma=15$~MeV.
\end{itemize}

For general filter design and performance comparisons, signal data from 1,820 windows (with 55 bunches each) were considered, forming a data stream with 100,100 samples. The simulation data were split into ten subsets of ten thousand and ten events each, in order to evaluate the statistical fluctuations. The error bars were computed through a cross-validation procedure~\cite{Duda2001}. In this procedure, one subset was kept for testing and did not participate in the filter design, which used the remaining nine subsets. The testing subset was shifted for each fold, so that each subset could play once the role of the testing subset along the cross-validation procedure. The average performance and the Root Mean Square (RMS) value from the ten models developed along such a procedure were computed as the estimate of the expected performance and the corresponding error bar, respectively.

\subsubsection{FIR Filter Design}

The baseline implementation was proposed in~\cite{Duarte2019} (see Figure~\ref{fig:fir}). For filter design, two parameters must be tuned: the filter order and the threshold value for reducing the estimation error.

\begin{figure}[htb]
\centerline{\includegraphics[width=.9\textwidth]{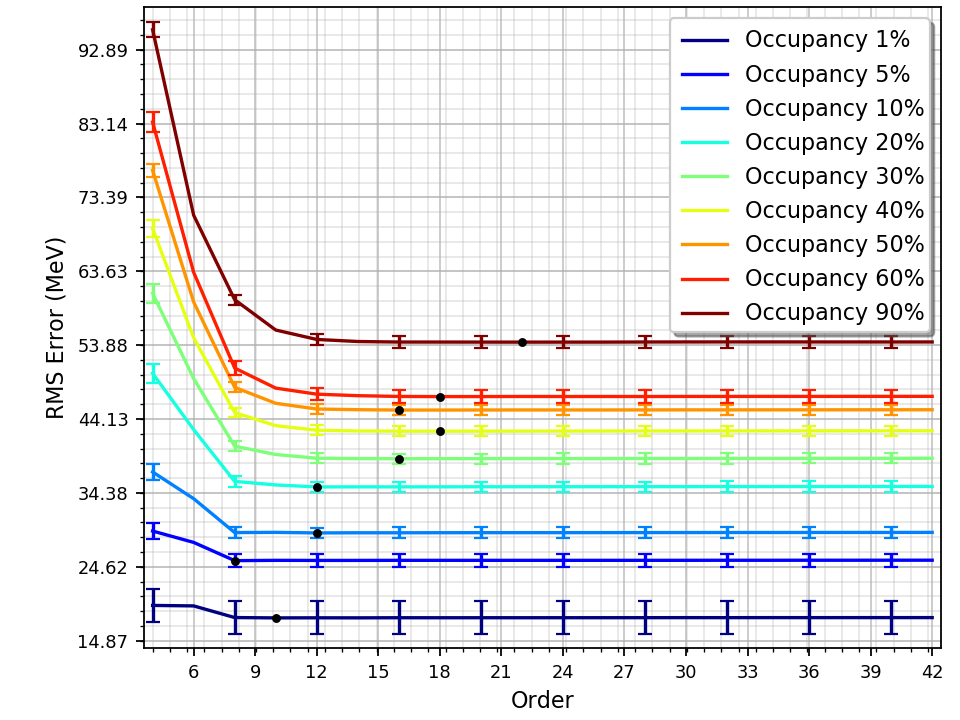}}
\caption{The estimation error as a function of the FIR filter order (threshold set as zero). The black dots point out the FIR Filter order values for which the minimum RMS errors were achieved.}
\label{fig:fir_order}
\end{figure}

\begin{figure}[htb]
\centerline{\includegraphics[width=.9\textwidth]{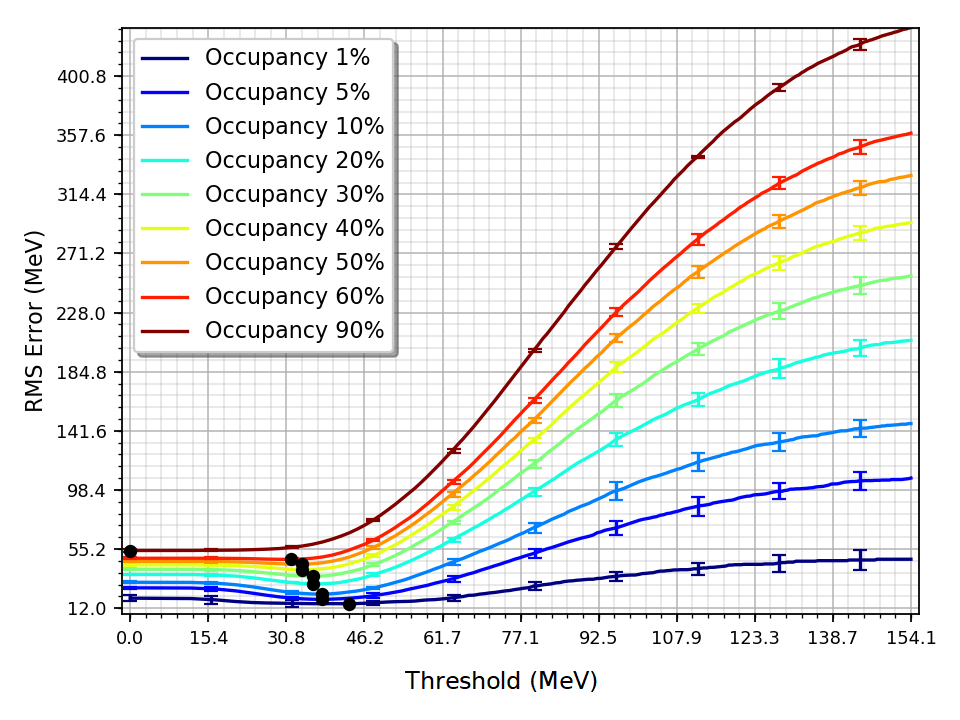}}
\caption{The estimation error as a function of the threshold value determined in the output of the FIR filter. The black dots point out the threshold values for which the minimum RMS errors were achieved.}
\label{fig:fir_order_rms}
\end{figure}

The energy estimation performance as a function of the filter order is shown in Figure~\ref{fig:fir_order}. In this figure, the output threshold was set to null aiming at avoiding negative energy estimations. Based on this figure, a filter order of~22 was kept fixed for further analysis (hence, the filter delay was $\Delta=11$ samples), regardless of the occupancy value, since the RMS error did not decrease for higher filter order values. For optimal threshold value determination, the estimation error as a function of the threshold value is plotted in Figure~\ref{fig:fir_order_rms} for each occupancy value. One can conclude that a threshold around 30~MeV might be used regardless of the occupancy level. As expected, the estimation error increased with the pile-up level.

\subsubsection{Matching Pursuit Parameters}

\begin{figure}[htb]
\centering
\begin{subfigure}{0.49\textwidth} \label{fig_first_case}
  \centering
  \includegraphics[width=\textwidth]{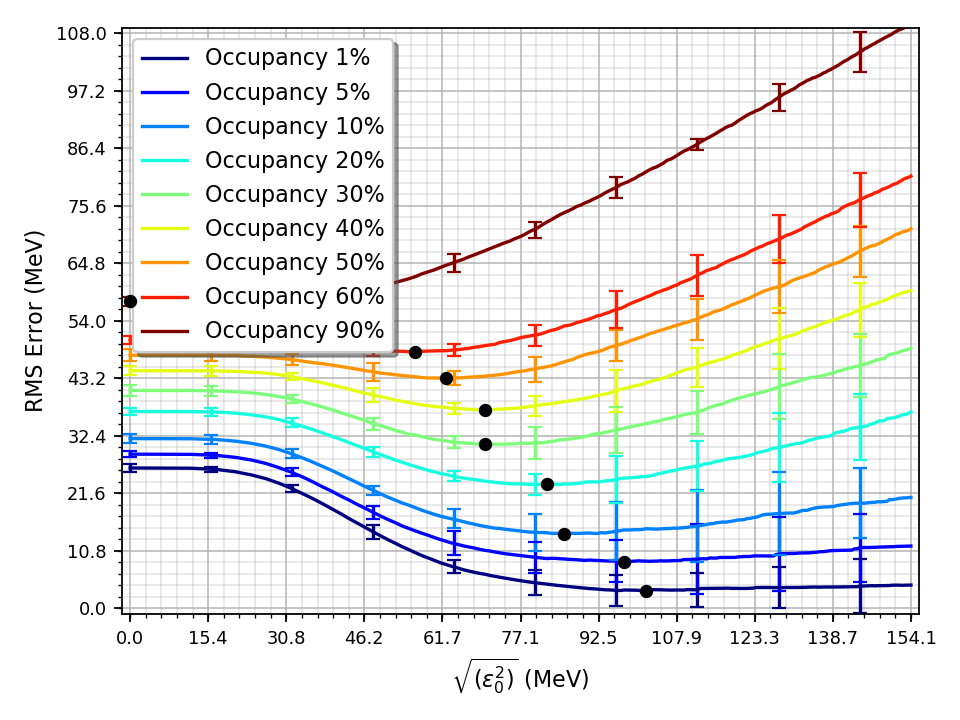}{}
  \caption{}
    \end{subfigure}%
    ~
    \begin{subfigure}{0.49\textwidth} \label{fig_second_case}
  \centering
  \includegraphics[width=\textwidth]{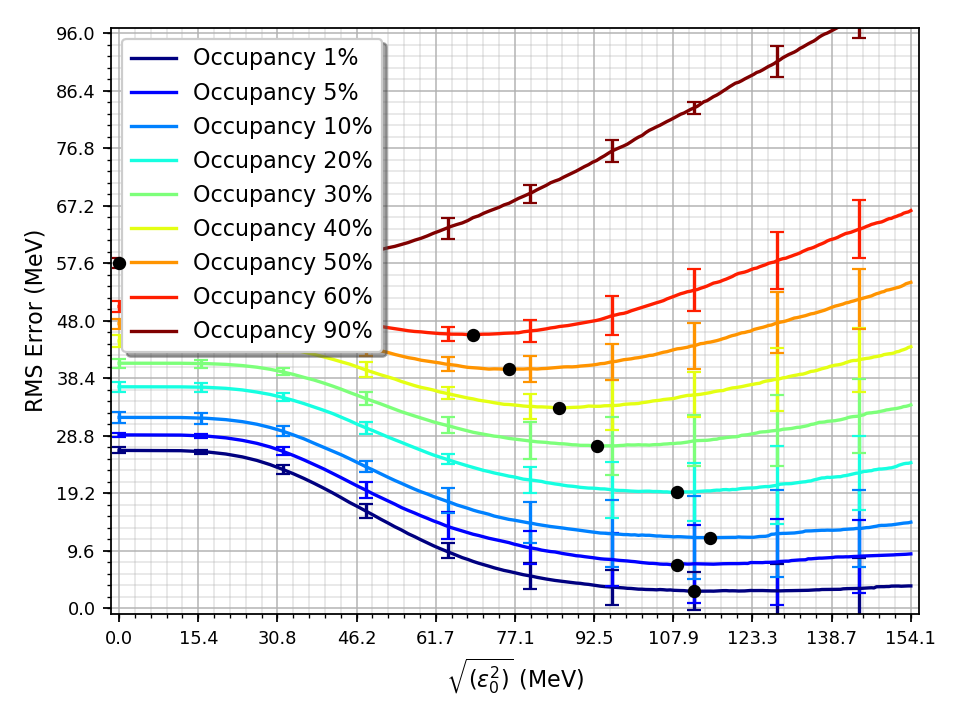}{}
  \caption{}
    \end{subfigure}
\caption{The RMS error as a function of the residue energy for both OMP (a) and LS-OMP (b) methods. The black dots point out the threshold values for which the minimum RMS errors were achieved.}
\label{fig:mp_rms}
\end{figure}

The unique parameter to be determined for the Matching Pursuit methods is the squared value for the residue limit ($\epsilon_0^2$). Figure~\ref{fig:mp_rms} shows such error limits as a function of this parameter for both the OMP and the LS-OMP methods. Since this parameter is measured in MeV$^{2}$,  its squared root value was employed for the sake of physical interpretation in terms of energy values in~MeV. As expected, the optimal residue value decreased as the occupancy level was increased, requiring more signals in the support matrix $\mathbf{H}_s$. Although both methods achieved similar performance, it is worth mentioning that the amount of operations for LS-OMP implementation is much higher, as it executes a complete estimation with Least Square for each iteration whereas the OMP method applies the linear regression only at the final stage.

\subsubsection{Separable Surrogate Functional Parameters}

For Separable Surrogate Functional, two parameters require tuning: the step-size~$\mu$ for the Coordinate-Descent iterations and the~$\lambda$ value for the shrinkage function. For faster convergence, the~$\mu$ value may be determined dynamically, depending on the current residue~$\mathbf{d}$. Since the argument in Equation~\ref{eq:shrink} is a linear Gradient-Descent iteration step, the optimal step-size~$\mu$ for the $i$-th iteration is~\cite{Bertsekas1999}

\begin{equation}
\mu = \frac{\mathbf{d}^{iT}\mathbf{d}^i}{\mathbf{d}^{iT}(\mathbf{H}^T\mathbf{H})\mathbf{d}^i}
\label{eq:mu}
\end{equation}
where $\mathbf{d}^i = (\mathbf{H}\mathbf{x}^i - \mathbf{r}$) is the residue vector at the $i$-th iteration.

\begin{figure}[htb]
\centerline{\includegraphics[width=.9\textwidth]{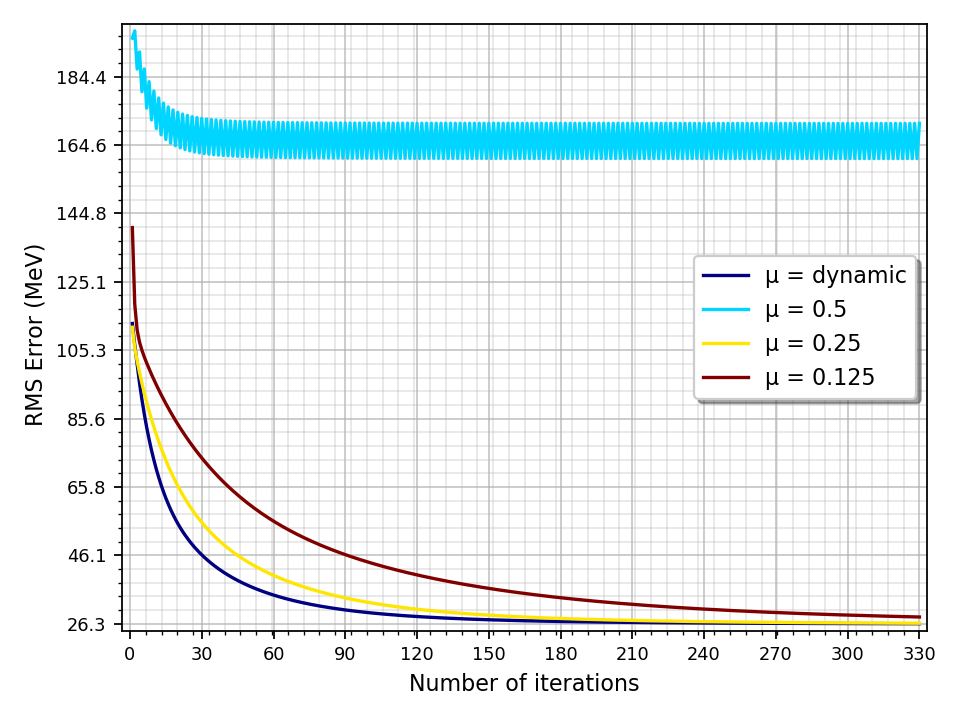}}
\caption{The RMS error as a function of the number of iterations. A cell with 30\% of occupancy level was employed.}
\label{fig:ssf_mu_mean}
\end{figure}

Figure~\ref{fig:ssf_mu_mean} shows the evolution of the RMS error as a function of the iteration index, for four different~$\mu$ values, when the occupancy level was kept fixed at 30\%. When evaluating the dependency with respect to the parameter~$\mu$, the other parameter ($\lambda$)  was set to zero. For these plots, the entire set (100,100 samples) was used to increase the statistics and no error bar is shown.

\begin{figure}[htb]
\centerline{\includegraphics[width=.9\textwidth]{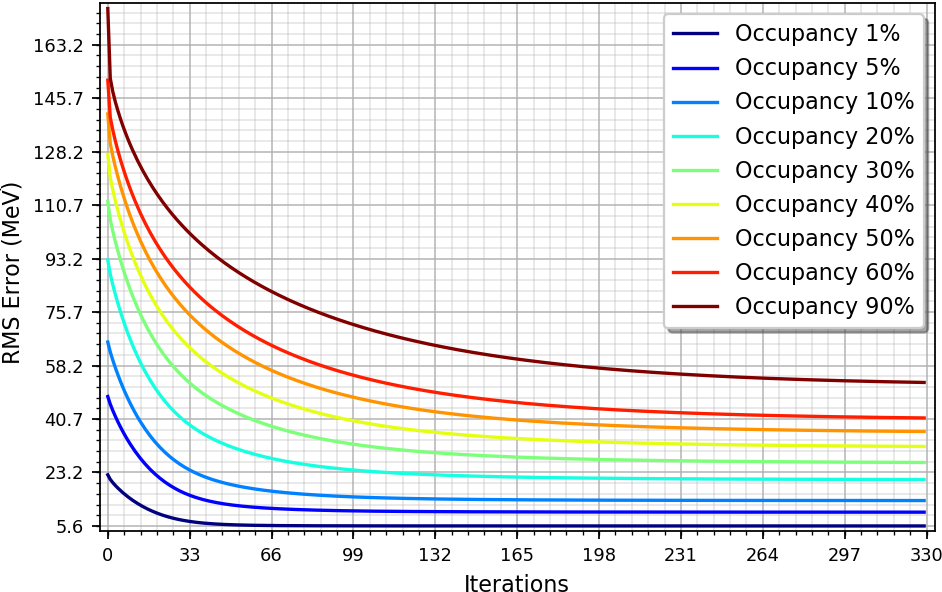}}
\caption{The RMS error as a function of the number of Separable Surrogate Functional iterations for several occupancy levels and $\mu=0.25$.}
\label{fig:fir_mu_025}
\end{figure}

The four evaluated values for~$\mu$ were: the dynamic one (according to Equation~\ref{eq:mu}) and three fixed values, in order to avoid the computation of Equation~\ref{eq:mu} in hardware. The three fixed values were chosen within the range of the values obtained through the dynamic approach, so that no multiplication circuitry would be necessary in a fixed-point numerical representation for any of these values in hardware (bit-shifting operation only)~\cite{Milne2016}. One may notice that $\mu=0.5$ was too high and the algorithm turned out to be unstable. On the other extreme, for $\mu=0.125$ the convergence was too slow. The best solution was found for $\mu=0.25$, which produced an evaluation performance similar to the dynamical one (see Figure~\ref{fig:ssf_mu_mean}). Figure~\ref{fig:fir_mu_025} summarises the performance for the Separable Surrogate Functional method as a function of the number of iterations for different occupancy levels, when $\mu=0.25$ was kept fixed.

\begin{figure}[htb]
\centerline{\includegraphics[width=.9\textwidth]{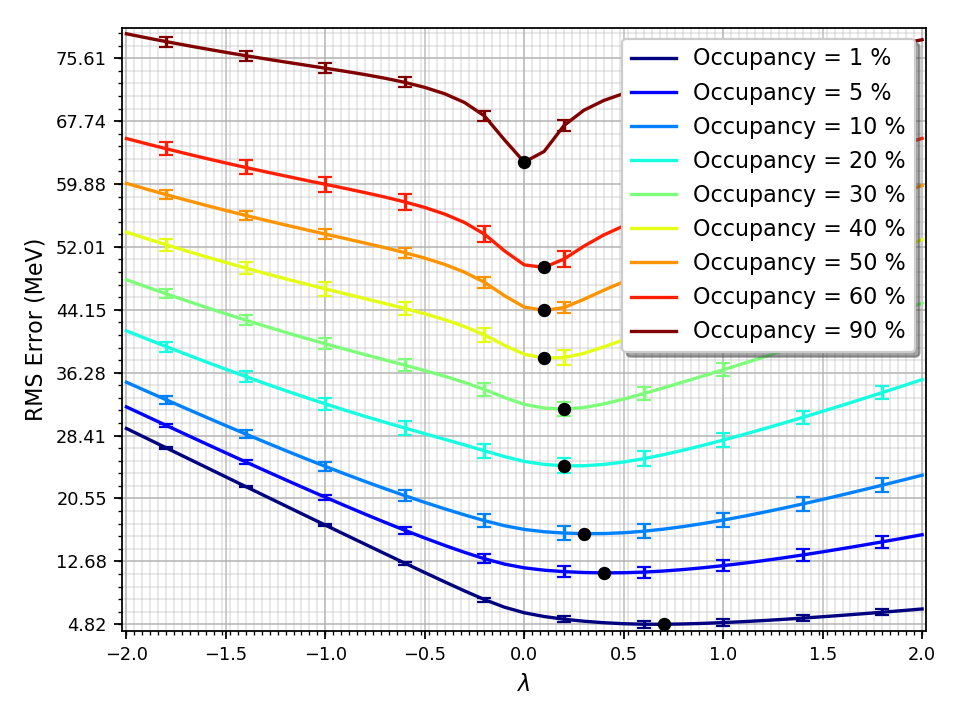}}
\caption{The RMS error as a function of the parameter $\lambda$ in the Separable Surrogate Functional method, considering several occupancy levels and $\mu=0.25$. A total of 330 iterations were applied in order to guarantee full convergence of the algorithm.  The black dots refer to the optimal values.}
\label{fig:fir_ssf_lambda}
\end{figure}

Concerning the tuning for the parameter~$\lambda$, Figure~\ref{fig:fir_ssf_lambda} shows how the estimation performance behaved around the optimal value for several occupancy levels. It can be seen that the value of the parameter~$\lambda$ tended to zero while the pile-up level increased. Therefore, for sake of hardware simplicity, the parameter~$\lambda$ was set to zero, regardless of the occupancy level.

\begin{figure}[htb]
\centerline{\includegraphics[width=.9\textwidth]{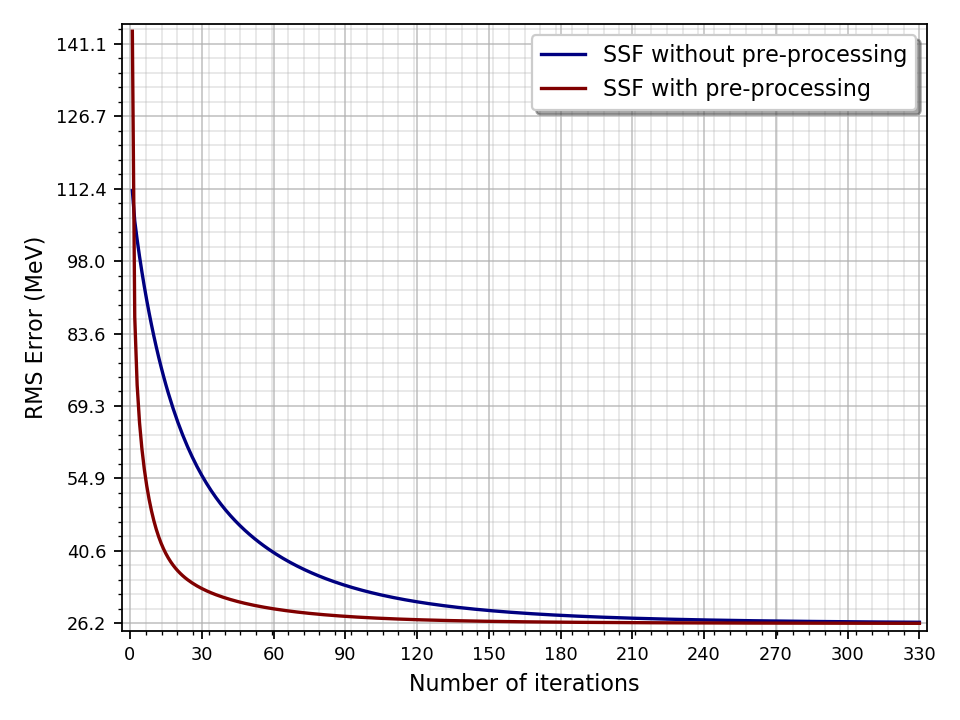}}
\caption{The RMS error as a function of the number of Separable Surrogate Functional iterations with and without the preprocessing step for $\mathbf{x}^0$ determination. A cell with $30\%$ of occupancy level was employed.}
\label{fig:fir_ssf_pseudo}
\end{figure}

The results in Figure~\ref{fig:fir_mu_025} were obtained by initializing the vector~$\mathbf{x}^0$ with the 48 central elements of the measured vector~$\mathbf{r}$. This was shown to be faster than initializing~$\mathbf{x}^0$ as a null vector. However, as can be seen from this figure, more than 200 iterations were still necessary for algorithm convergence for most of the occupancy levels. Thus, in this work, the initialization of~$\mathbf{x}$ is proposed to follow Equation~\ref{eq:x0} and Figure~\ref{fig:fir_ssf_pseudo} compares the evaluation of the estimation error as the number of iterations increases, with and without this preprocessing scheme, for a cell with~$30\%$ of occupancy level. One can notice that the convergence proved to be faster when the proposed initialization was applied.

\subsection{Performance}

Once the parameters for the aforementioned methods were all set to their design tuned values, a performance comparison was carried out. Table~\ref{tab:rms} summarizes the final performance. In this table, Linear Programming (as described in~\cite{Barbosa2017}) corresponds to a direct implementation of the problem $P_{1,\epsilon_{0}}$ (Equation~\ref{eq:p1}), and FIR corresponds to the method proposed in~\cite{Duarte2019}. The last three columns refer to the methods proposed in this paper.

\begin{table}[htbp]
\centering
\caption{Performance estimation in terms of the RMS error values (in MeV). Best performance for each occupancy level is shown in boldface.}
\label{tab:rms}
\begin{tabular}{ccccccccc}
\hline
\textbf{\begin{tabular}[c]{@{}c@{}}Occupancy\\ level\end{tabular}} & \textbf{\begin{tabular}[c]{@{}c@{}}Linear\\ Programming\end{tabular}} & \textbf{\begin{tabular}[c]{@{}c@{}}FIR\\ Filter\end{tabular}} & \textbf{OMP} & \textbf{LS-OMP} & \textbf{SSF}$^*$ \\ \hline
1 \%  &   4.44$\pm1.07$  & 15.22$\pm2.33$  &  3.27$\pm0.75$   & \textbf{ 2.86$\pm0.49$} &          3.72$\pm0.33$  \tabularnewline \hline
5 \%  &  10.13$\pm0.67$  & 18.48$\pm0.91$  &  8.77$\pm0.56$   & \textbf{ 7.29$\pm0.35$} &          9.21$\pm0.62$  \tabularnewline \hline
10 \% &  14.9$\pm0.93$   & 22.3$\pm0.74$   & 13.95$\pm0.9 $   & \textbf{11.8$\pm0.91$}  &         13.28$\pm0.84$  \tabularnewline \hline
20 \% &  22.12$\pm0.83$  & 29.76$\pm0.68$  & 23.27$\pm0.64$   & \textbf{19.36$\pm0.79$} &         20.51$\pm0.92$  \tabularnewline \hline
30 \% &  28.47$\pm0.9 $  & 35.33$\pm0.69$  & 30.79$\pm1.03$   &         27.12$\pm0.73$  & \textbf{26.26$\pm0.82$} \tabularnewline \hline
40 \% &  34.32$\pm0.92$  & 40.27$\pm0.65$  & 37.26$\pm0.86$   &         33.45$\pm0.93$  & \textbf{31.67$\pm0.9 $} \tabularnewline \hline
50 \% &  39.03$\pm0.91$  & 44.12$\pm0.7 $  & 43.25$\pm1.09$   &         39.91$\pm0.83$  & \textbf{36.57$\pm0.68$} \tabularnewline \hline
60 \% &  45.6$\pm0.97$   & 47.72$\pm0.8 $  & 48.17$\pm0.77$   &         45.67$\pm0.97$  & \textbf{41.01$\pm0.98$} \tabularnewline \hline
90 \% &  59.21$\pm0.87$  & 54.15$\pm0.76$  & 57.68$\pm0.88$   &         57.68$\pm0.88$  & \textbf{52.63$\pm0.65$} \tabularnewline \hline
\end{tabular}\\
$^*$Separable Surrogate Functional (SSF) parameters set as $\mu=0.25$ and $\lambda=0$
\end{table}

For lower occupancy levels, the target sequence ($x[n]$) presented high sparsity and the Sparse Representation methods outperformed the FIR filter considerably. As the occupancy level was increased, the sparsity assumption gradually lost its strength. Therefore, the performance of the Sparse Representation methods got closer to the one from FIR filters for occupancy values above 30\%. Among the Matching Pursuit methods, LS-OMP was slightly better. However, the Matching Pursuit methods were less efficient with respect to Separable Surrogate Functional for higher occupancy levels. The first step of the Matching Pursuit algorithms, the vector support determination, presented lower efficiency when the sparsity of the target sequence was not so high. Concerning the offline method used for comparison (Linear Programming), the achieved results were similar to the ones from LS-OMP. 

In general, the occupancy levels shown in Table~\ref{tab:rms} may be considered practical for different designs in modern collider’s experiments like ATLAS and CMS, in LHC, and for technologies which are being considered for future colliders. Apart from the nominal collider luminosity, the occupancy level is also strongly correlated to the detector geometry parameters, such as pseudo-rapidity and radial layer. For forward calorimeters (with very high pseudo-rapidity), occupancy values at 90\% level are already found in the LHC detectors, while in low pseudo-rapidity regions, the occupancy near 1\% is the one mostly found. However, as the luminosity of the colliders is foreseen to increase continuously, the necessity of using methods more resilient to pile-up effects, such the one proposed here, becomes more evident. In the Run 3, for example, the luminosity is expected to increase considerably.

\subsubsection{Detection Efficiency}

\begin{figure}[htp]
\centering
\begin{subfigure}{0.49\textwidth}
  \centering
  \includegraphics[width=\textwidth]{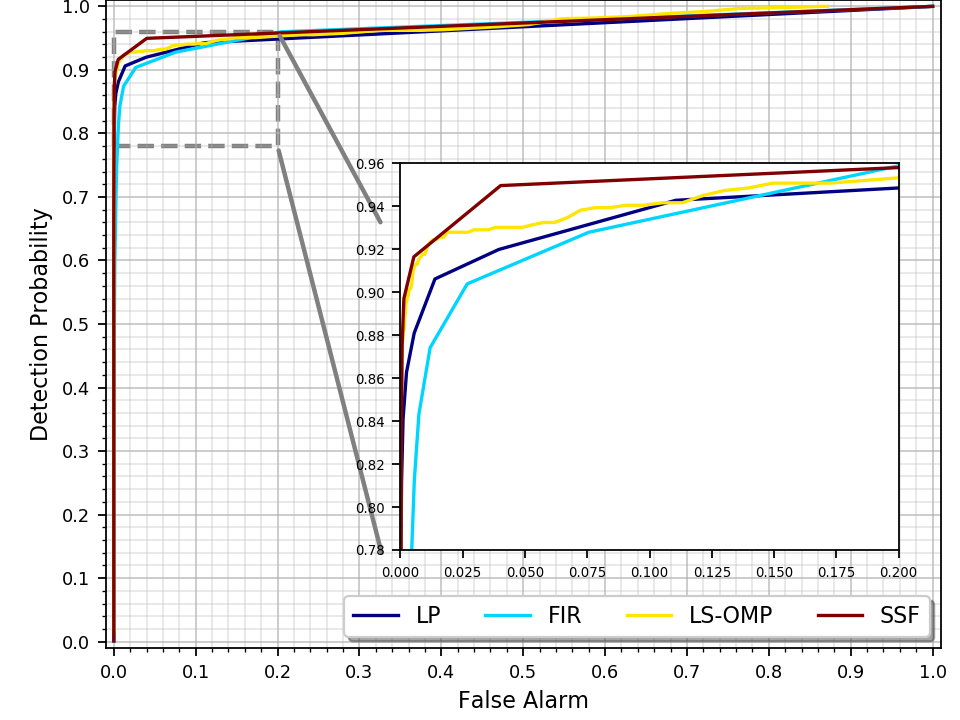}{}
  \caption{1\% of occupancy}
    \end{subfigure}%
    ~
    \begin{subfigure}{0.49\textwidth}
  \centering
  \includegraphics[width=\textwidth]{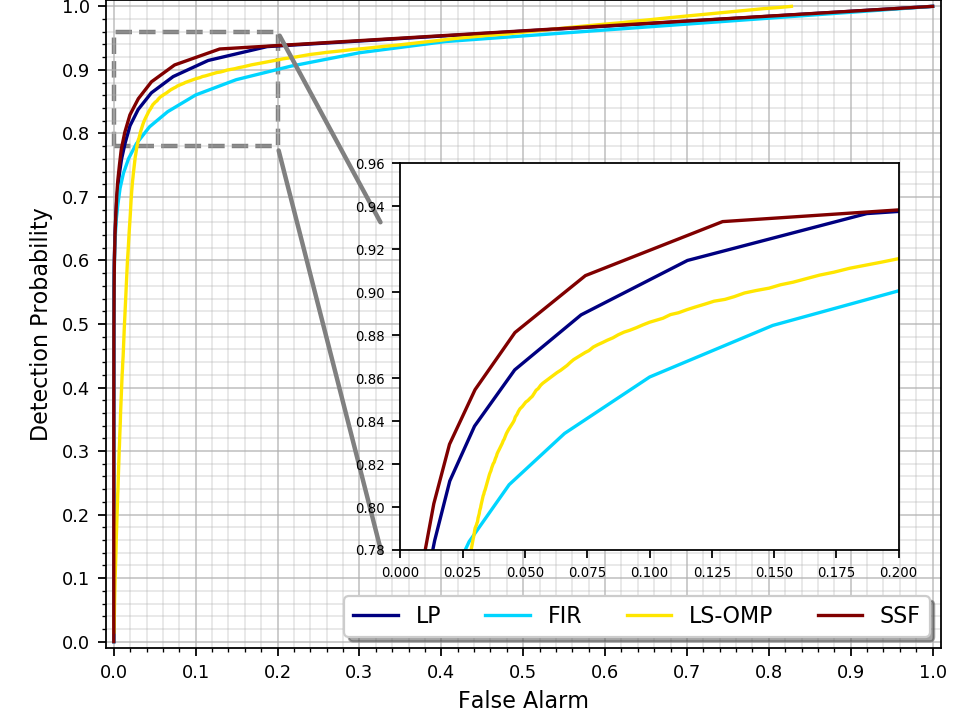}{}
  \caption{30\% of occupancy}
    \end{subfigure}
\caption{The ROC curves for low (a) and high (b) occupancy values.}
\label{fig:roc}
\end{figure}

In this section, performance evaluation in terms of detection efficiency is outlined. The analysis was performed through the Receiver Operating Characteristic (ROC) curve~\cite{Fawcett2006}, which shows how the Probability of Detection (PD) for a given signal varies when a detection threshold is applied in the energy estimation value and the corresponding impact in the False Alarm (FA) probability. Here, PD comprised the percentage of bunches crossings with hits that were correctly detected (estimated energy above the threshold) and FA was the percentage of bunches crossings with no hits that have been wrongly selected. Figure~\ref{fig:roc} presents the ROC curves for two different occupancy levels (1\% and 30\%). The two tested Matching Pursuit methods (OMP and LS-OMP) presented similar results. Thus, only LS-OMP is shown in this analysis. For 1\% of occupancy, both FIR filter and Separable Surrogate Functional methods presented similar detection efficiency. As the occupancy was increased, the advantage of the Sparse Representation methods, when compared to the FIR filter, became more evident. The Separable Surrogate Functional method remained as one with the highest detection efficiency, regardless of the occupancy level (1\% or 30\%).

\subsubsection{Analysis Summary}

From the above analysis, one can conclude that, for lower occupancy conditions, if only the detection efficiency is taken into account (triggering purpose only), the FIR filter structure is a competitive choice, as it can be seen from the ROC curves. On the other hand, if the estimation accuracy is also an important issue, Matching Pursuit methods may be employed. For high pile-up levels, the Separable Surrogate Functional method may be chosen when both estimation accuracy and detection efficiency are of main concern. Besides, even for lower occupancy values, the Separable Surrogate Functional method was competitive with respect to general energy estimation performance.

The Separable Surrogate Functional method has also the lowest computational cost among the Sparse Representation one. Therefore, the next section evaluates an online implementation of the Separable Surrogate Functional method in FPGA technology and does a comparison with the FIR filter implementation in terms of technology resource requirements.

\section{Hardware Implementation and Cost Analysis}\label{sec:hardware}

Aiming at optimizing the hardware resource utilization, the argument in Equation~\ref{eq:shrink} is rewritten according to:

\begin{equation}
    \mathbf{x}^{i} + \mu(\mathbf{H}^T\mathbf{r} - \mathbf{H}^T\mathbf{H}\mathbf{x}^i)
\label{eq:it}
\end{equation}
where $\mathbf{H}^T\mathbf{H}$ is a constant Toeplitz matrix~\cite{Bottcher2000}, which is computed offline. Concerning logic circuitry allocation, the Toeplitz matrix presents a certain symmetry that allows to reduce the cost through parallel implementation. In the text and in the figures that follow, this matrix is identified as~$\mathbf{Y}$ for sake of simplicity. For the same reason, the vector~$\mathbf{H}^T\mathbf{r}$ is also previously computed in hardware, just before the first iteration, and it will be indicated as a constant vector~$\mathbf{f}$.

Other important simplification concerns the non-linearity imposed by the shrinkage function (Figure~\ref{fig:shrink}). For $\lambda=0$, this function is implemented by a simple logic comparison: if the value is lower than zero, the component of the vector~$\mathbf{x}$ is replaced by zero before the next iteration.

Figure~\ref{fig:block} shows the block diagram for Separable Surrogate Functional algorithm implementation. The blocks in gray color execute matrix computations altogether, in parallel (distributed combinational circuitry) and in fixed-point arithmetic~\cite{Padgett2009}. In this figure, it is possible to identify a system with three pipeline stages~\cite{Oda1996}, split into input and output register banks, where:

\begin{figure}[htb]
\centering
\includegraphics[width=.45\textwidth]{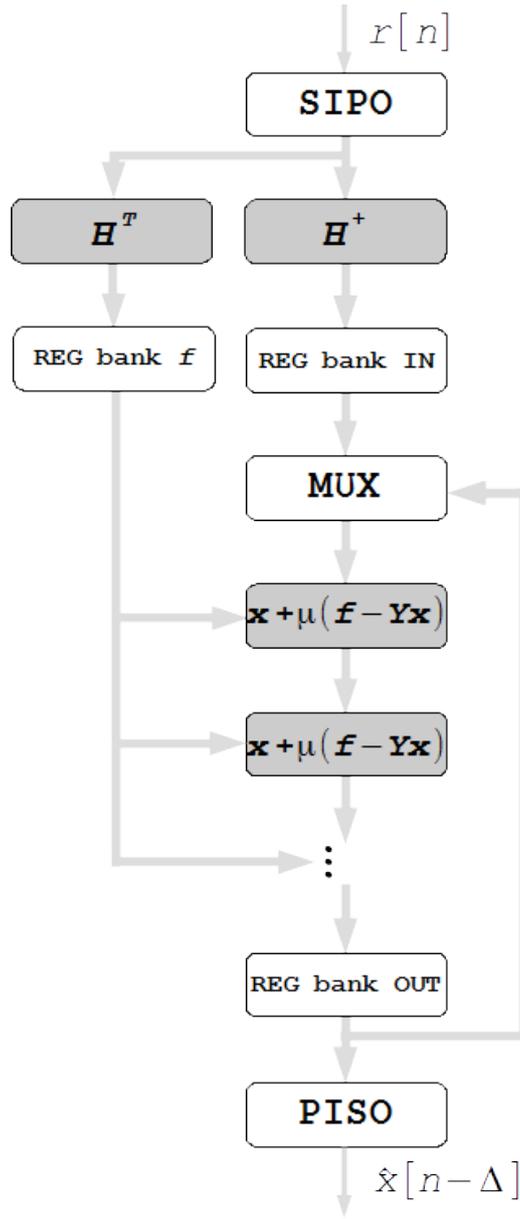}
\caption{The block diagram of the proposed circuitry. It operates synchronously with the $r[n]$ input signal.}
\label{fig:block}
\end{figure}

\begin{enumerate}

\item The first pipeline stage comprises the SIPO (Serial-Input Parallel-Output) circuitry, the block which computes the $\mathbf{f} = \mathbf{H}^T\mathbf{r}$ operation and the preprocessing for $\mathbf{x}^0$ computation. The SIPO assemblies a window of 55~samples (the vector~$\mathbf{r}$) buffering the most recent consecutive $r[n]$ samples. When the SIPO internal registers are fully occupied, both vectors~$\mathbf{f}$ and~$\mathbf{x}^0$ are computed. In the next clock rising edge, the 48 resulting components, for both vectors, are stored in input register banks. This process is repeated to each window (55~clock cycles) uninterruptedly.

\item The second pipeline stage comprises the blocks between the input and the output register banks. The ones in gray color are responsible to execute one Separable Surrogate Functional iteration. This stage has a feedback path, synchronous to the data clock (40~MHz, if LHC would be the target application), in order to compute several iterations before the next window is delivered by the first pipeline stage. The resulting reconstructed vector ($\mathbf{x}^i$) is updated in the output register bank to each clock. 

\item The third and last stage is executed by the PISO (Parallel-Input Serial-Output) block. This circuitry serialises the reconstructed vector present in the output register bank, delivering the free-running $\hat{x}[n-\Delta]$ information, synchronous with the ADC clock~(40~MHz). In this way, the windowed processing of the proposed system becomes transparent.

\end{enumerate}

\subsection{Firmware Implementation}

As described before, the system operates in a three stage pipeline architecture. Thus, three consecutive windows are evaluated simultaneously, one at each stage. The first stage buffers the most recent window and sends the data, in parallel, to the next stage.

The Separable Surrogate Functional iterations are implemented by the second pipeline stage. In the first clock, the multiplexer is switched to the input register. In this way, the initial value of the algorithm is $\mathbf{x}^0$, as expected. This signal propagates through the blocks that compute one Separable Surrogate Functional iteration. For the consecutive 55 clock cycles, the multiplexer points to the output register, which updates the reconstructed vector, in a feedback data-flow, at each clock cycle. In this stage, if only one Separable Surrogate Functional iteration block is implemented, it is possible to compute 55~iterations before the first stage sends a new fresh window. Therefore, a cascade of $z$ blocks allows the execution of $55\times z$ iterations.

The last stage simply serializes the most recent reconstructed window delivered by the second stage. As a consequence of this proposed architecture, the response delay $\Delta$ is kept fixed in 2 times the window size ($2 \times 55 = 110$), regardless the number of iterations (which is a multiple of 55).

\subsection{Cost Analysis}

The proposed circuitry was implemented in FPGA. The Software Vivado~\cite{Sanjay2016} was used in the design, and Verilog~\cite{Monk2017} was the Hardware Description Language (HDL)~\cite{Botros2015}. For performance tests, the chip XCZU7EV-FFVC1156-2-E from the Zynq UltraScale+ family~\cite{Zynq2019} was chosen, since its evaluation is free and can be crosschecked without costs. For representation, 17~bits (7 bits for range and 10 bits for precision) were selected for fixed-point arithmetic, for all the parts of the circuitry. This configuration was chosen empirically after exhaustive simulation with several different numerical representations. Table~\ref{tab:fpga} summarizes the implementation performance. As for comparison, the results for a FIR filter structure are also outlined, and the details about the hardware implementation of the FIR filter may be found in~\cite{Duarte2019}. In this table, it is shown the performance results for up to three cascaded iteration blocks. For more than this, the operation frequency is below 40~MHz. Among the Separable Surrogate Functional implementations, the one with only one iteration block together with the preprocessing circuit is an attractive solution in terms of cost-efficiency.

\begin{table}[htbp]
\caption{Implementation resources in terms of different hardware elements and frequency operation.}
\label{tab:fpga}
\centering
\begin{tabular}{|l|l|l|l|l|l|l|l|}
\hline
\multicolumn{1}{|c|}{\multirow{2}{*}{Metrics}} & \multicolumn{1}{c|}{\multirow{2}{*}{FIR}} & \multicolumn{2}{c|}{SSF 1 Block} & \multicolumn{2}{c|}{SSF 2 Blocks} & \multicolumn{2}{c|}{SSF 3 Blocks} \\ \cline{3-8} 
\multicolumn{1}{|c|}{} & \multicolumn{1}{c|}{} & \textbf{Pp} & \textbf{nPp} & \textbf{Pp} & \textbf{nPp} & \textbf{Pp} & \textbf{nPp} \\ \hline
LUT** & 1,447 & 31,378 & 33,132 & 59,927 & 63,966 & 87,160 & 94,870 \\ \hline
FF** & 731 & 2,097 & 2,625 & 2,098 & 2,625 & 2,097 & 2,625 \\ \hline
DSP** & 125 & 435 & 308 & 503 & 376 & 571 & 444 \\ \hline
Freq (MHz) & 56.76 & 71.20 & 79.17 & 50.34 & 57.39 & 41.47 & 42.22 \\ \hline
RMS (MeV) & 40 & 31 & 43 & 29 & 32 & 28 & 29 \\ \hline
\end{tabular}\\
*\textit{Pp = with preprocessing  nPp = without preprocessing} \\
**\textit{Number of: LUT = Look-up-tables, FF = Flip-Flops, DSP = Digital Signal Processor}
\end{table}

As it can be seen from previous analysis, the Separable Surrogate Functional implementation requires $15\times$ more resources, on average, than FIR filter structure. However, the proposed method presents a higher estimation performance, which may justify its implementation, depending on the available resources. It is worth to mention that the amount of resources for Separable Surrogate Functional implementation is proportional to the square of the window length. The value of 55 used here is just one example of direct implementation. Ongoing analyses are investigating whether splitting the processing window in several smaller windows may improve the implementation requirements. For practical considerations, Table~\ref{tab:cost} presents the resources used to fit the Separable Surrogate Functional method in three different FPGA technologies, where the feasibility of
implementation becomes evident.

\begin{table}[htbp]
\caption{Resources spent in commercial FPGA devices.}
\label{tab:cost}
\centering
\begin{tabular}{|l|l|l|l|l|}
\hline
\multicolumn{1}{|c|}{\multirow{2}{*}{Metrics}} & \multicolumn{1}{c|}{SSF 3 Blocks$^1$} & \multicolumn{1}{c|}{XCZU7EV$^2$} & \multicolumn{1}{c|}{XC7VX1140T$^3$} & \multicolumn{1}{c|}{XCVU440$^4$} \\ \cline{2-5} 
\multicolumn{1}{|c|}{} & \multicolumn{1}{c|}{\textbf{nPp}} & \multicolumn{1}{c|}{\textbf{low cost}} & \multicolumn{1}{c|}{\textbf{middle cost}} & \multicolumn{1}{c|}{\textbf{high cost}} \\ \hline
LUT & 94,870 & 41.17\% (230,400) & 13,32\% (712,000) & 3,74\% (2,532,960) \\ \hline
FF & 2,625 & 0.56\% (460,800) & 0,18\% (1,424,000) & 0,05\% (5,065,920) \\ \hline
DSP & 444 & 25.69\% (1,728) & 13,21\% (3,360) & 15,41\% (2,880) \\ \hline
\end{tabular}\\
$^1$\textit{SSF 3 Blocks without preprocessing} \\
$^2$\textit{XCZU7EV-FFVC1156-2-E} $^3$\textit{XC7VX1140T-1FLG1930C-ND} $^4$\textit{XCVU440-2FLGA2892E}
\end{table}

\section{Conclusion}

This paper presented an iterative method for a matrix based deconvolution process, which can be implemented for online operation. The proposed method is based on established sparse theory, over-performing the previously proposed FIR filter based deconvolution method. The method based on Iterative Shrinkage has shown to be a good choice for implementation and a FPGA circuitry was proposed. Compared with a simpler FIR filter design, despite the increase in cost implementation, it is feasible when modern FPGA technology is considered. The gain in estimation accuracy is evident, with improvement between 3\% and 300\% (depending on the occupancy level), when compared with FIR filter implementation. Therefore, it may justify its usage in experiments with intense pile-up levels.

\section*{Acknowledgment}

The authors would like to thank CNPq, CAPES, RENAFAE-MCTI, FAPERJ,
FAPEMIG (Brazil) and CERN (Switzerland) for their support to this work.

\bibliographystyle{JHEP}
\bibliography{main}
\end{document}